\documentclass{acm_proc_article-sp}

\usepackage{algorithm}
\usepackage{algpseudocode}
\usepackage{amsbsy}
\usepackage{color}
\usepackage{enumerate}
\usepackage{float}
\usepackage{graphicx}
\usepackage{hyperref}
\usepackage{breakurl}
\usepackage[final]{listings}
\usepackage{multirow}
\usepackage[caption=false]{subfig}
\usepackage{tikz}
\usepackage{verbatim}
\usepackage{xspace}

\newcommand{\ie}{{\em i.e.}\xspace}

\newcommand{\Figref}[1]{Figure~\ref{#1}}

\algnewcommand\algorithmicswitch{\textbf{switch}}
\algnewcommand\algorithmiccase{\textbf{case}}
\algnewcommand\algorithmicassert{\texttt{assert}}
\algnewcommand\Assert[1]{\State \algorithmicassert(#1)}%
\algdef{SE}[SWITCH]{Switch}{EndSwitch}[1]{\algorithmicswitch\ #1\ \algorithmicdo}{\algorithmicend\ \algorithmicswitch}%
\algdef{SE}[CASE]{Case}{EndCase}[1]{\algorithmiccase\ #1}{\algorithmicend\ \algorithmiccase}%
\algtext*{EndSwitch}%
\algtext*{EndCase}%

\newcommand{\PRAGMATIC}{PRAgMaTIc\xspace}

\definecolor{listinggray}{gray}{0.9}
\definecolor{lbcolor}{rgb}{0.9,0.9,0.9}
\floatstyle{plain}
\newfloat{code}{h}{}
\floatname{code}{Code Snippet}

\begin{document}

\title{An Interrupt-Driven Work-Sharing For-Loop Scheduler}

\numberofauthors{3}
\author{
\alignauthor
Georgios Rokos\\
       \affaddr{Department of Computing,}\\
       \affaddr{Imperial College London,}\\
       \affaddr{London SW7 2AZ, UK}\\
       \email{gr409@doc.ic.ac.uk}
\alignauthor
Gerard J. Gorman\\
       \affaddr{Department of Earth Science and Engineering,}\\
       \affaddr{Imperial College London,}\\
       \affaddr{London SW7 2AZ, UK}\\
       \email{g.gorman@imperial.ac.uk}
\alignauthor
Paul H. J. Kelly\\
       \affaddr{Department of Computing,}\\
       \affaddr{Imperial College London,}\\
       \affaddr{London SW7 2AZ, UK}\\
       \email{p.kelly@imperial.ac.uk}
}

\maketitle

\begin{abstract}

In this paper we present a parallel for-loop scheduler which is based on 
work-stealing principles but runs under a completely cooperative scheme. POSIX 
signals are used by idle threads to interrupt left-behind workers, which in 
turn decide what portion of their workload can be given to the requester. We 
call this scheme Interrupt-Driven Work-Sharing (IDWS). This article describes 
how IDWS works, how it can be integrated into any POSIX-compliant OpenMP 
implementation and how a user can manually replace OpenMP parallel for-loops 
with IDWS in existing POSIX-compliant C++ applications. Additionally, we
measure its performance using both a synthetic benchmark with varying
distributions of workload across the iteration space and a real-life
application on Sandy Bridge and Xeon Phi systems. Regardless the workload
distribution and the underlying hardware, IDWS is always the best or among the
best-performing strategies, providing a good all-around solution to the
scheduling-choice dilemma.
\end{abstract}


\section{Introduction}
\label{sect:intro}
Most parallelism in shared-memory parallel programming comes from loops of
independent iterations, \ie iterations which can be safely executed in parallel.
However, distributing the iteration space over the available computational
resources of a system is not always a simple thing. Fine-grained control of
distribution is often associated with high overhead whereas static partitioning
of the iteration space can lead to significant load imbalance. In both cases,
the impact on performance is serious.

Research on an advanced for-loop scheduler was motivated by our work on 
\PRAGMATIC \cite{DBLP:journals/corr/RokosGSK13}, a hybrid OpenMP/MPI mesh
adaptivity framework. Profiling data revealed that many of \PRAGMATIC's parallel
loops are highly diverse, involving irregular computations which introduce high
levels of iteration-to-iteration load imbalance. Existing scheduling strategies
provided by OpenMP fail to achieve good balance with low scheduling overhead,
whereas adaptive mesh algorithms which constantly modify mesh topology make it
impossible to balance workload a priori.

We wanted the new scheduler to be portable and easily plug-able into the 
widely-adopted OpenMP API, so that it can target an as wide as possible range 
of systems, like Fujitsu's FX-10, a SPARC64-based supercomputer
\cite{fujitsu:fx10}. Those portability requirements prohibit the use of
platform- or vendor-specific threading mechanisms and parallel libraries, like
Intel\textsuperscript \textregistered Cilk\texttrademark Plus
\cite{Robison:2013:CPP:2498335.2498450,Frigo:1998:ICM:277652.277725,
Blumofe95cilk:an}. On the contrary, they call for a POSIX-compliant
implementation, based on the fact that most operating systems used in scientific
computing are POSIX-compliant and most compilers (e.g. Linux versions of gcc,
icc, xlc, etc.) implement OpenMP threads as POSIX threads (we have found it out
by experimenting with those compilers). Of course, since every OS has threading
and signalling mechanisms, the new scheduler can be implemented into any
compiler on any OS.

The main contributions of this article are the following:
\begin{itemize}
\item Present an new interrupt-driven work-sharing scheduler (IDWS) which can
easily be used with existing POSIX-compliant OpenMP applications
\item Demonstrate how OpenMP loops can be converted to IDWS loops
\item Describe how a compiler vendor can incorporate the new scheduler into
their product
\item Show using a variety of benchmarks that IDWS is a good all around solution
to the scheduling-choice dilemma, always being among the best-performing
strategies in all benchmarks
\end{itemize}

The rest of this paper is organized as follows: Section \ref{sect:background} 
provides an overview of loop scheduling options currently available, listing 
their advantages and weaknesses. In Section \ref{sect:new_alg} we describe the 
new scheduler, the way it works and explain why it offers a better tradeoff 
between load balance and scheduling overhead compared with other alternatives. 
Section \ref{sect:implementation} goes into details about the current 
implementation in C++ and explains how it can be used to replace OpenMP 
for-loops in existing codes. We demonstrate the scheduler's performance in 
Section \ref{sect:results} using both a synthetic benchmark and real for-loops 
from \PRAGMATIC. Finally, we conclude the paper in Section 
\ref{sect:conclusions}.

\section{Background}
\label{sect:background}

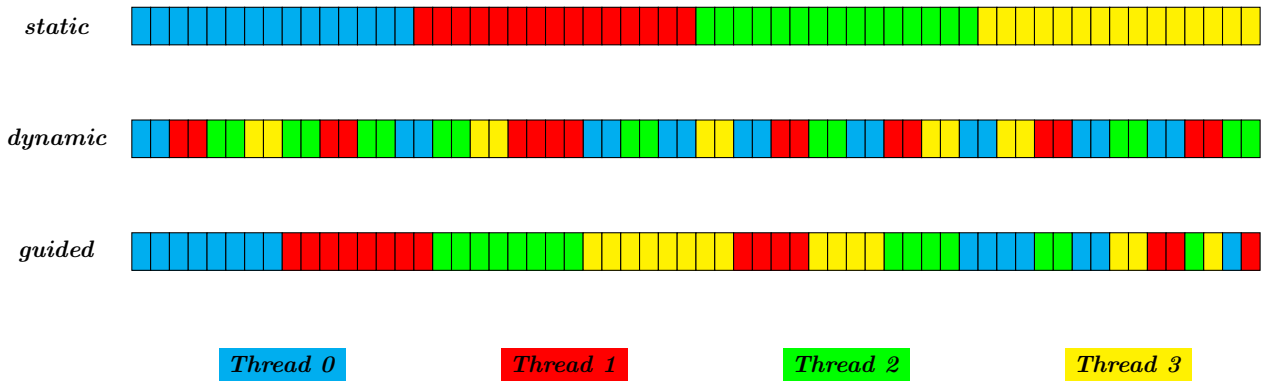
\begin{figure*}
\begin{center}
\begin{tikzpicture}
\node[draw=none,fill=none] at (0,3) {\textbf{\textit{static}}};
\foreach \colour / \offset in {cyan/1, red/4.75, green/8.5, yellow/12.25}
  \foreach \x in {0, 0.25, ..., 3.5}
    \draw[fill=\colour] (\x+\offset, 2.75) rectangle (\x+\offset+0.25, 3.25);

\node[draw=none,fill=none,align=center] at (0,1.5) {\textbf{\textit{dynamic}}};
\foreach \x in {0, 7, 12, 14, 16, 19, 22, 25, 27}
{
  \draw[fill=cyan] (1+\x*0.5, 1.25) rectangle (1.25+\x*0.5, 1.75);
  \draw[fill=cyan] (1.25+\x*0.5, 1.25) rectangle (1.5+\x*0.5, 1.75);
}
\foreach \x in {1, 5, 10, 11, 17, 20, 24, 28}
{
  \draw[fill=red] (1+\x*0.5, 1.25) rectangle (1.25+\x*0.5, 1.75);
  \draw[fill=red] (1.25+\x*0.5, 1.25) rectangle (1.5+\x*0.5, 1.75);
}
\foreach \x in {2, 4, 6, 8, 13, 18, 26, 29}
{
  \draw[fill=green] (1+\x*0.5, 1.25) rectangle (1.25+\x*0.5, 1.75);
  \draw[fill=green] (1.25+\x*0.5, 1.25) rectangle (1.5+\x*0.5, 1.75);
}
\foreach \x in {3, 9, 15, 21, 23}
{
  \draw[fill=yellow] (1+\x*0.5, 1.25) rectangle (1.25+\x*0.5, 1.75);
  \draw[fill=yellow] (1.25+\x*0.5, 1.25) rectangle (1.5+\x*0.5, 1.75);
}

\node[draw=none,fill=none] at (0,0) {\textbf{\textit{guided}}};
\foreach \colour / \block in {cyan/0, red/8, green/16, yellow/24}
{
  \pgfmathtruncatemacro \from {\block}
  \pgfmathtruncatemacro \to {\block+7}
  \foreach \x in {\from, ..., \to}
    \draw[fill=\colour] (1+\x*0.25, -0.25) rectangle (1.25+\x*0.25, 0.25);
}
\foreach \colour / \block in {cyan/44, red/32, green/40, yellow/36}
{
  \pgfmathtruncatemacro \from {\block}
  \pgfmathtruncatemacro \to {\block+3}
  \foreach \x in {\from, ..., \to}
    \draw[fill=\colour] (1+\x*0.25, -0.25) rectangle (1.25+\x*0.25, 0.25);
}
\foreach \colour / \block in {cyan/50, red/54, green/48, yellow/52}
{
  \pgfmathtruncatemacro \from {\block}
  \pgfmathtruncatemacro \to {\block+1}
  \foreach \x in {\from, ..., \to}
    \draw[fill=\colour] (1+\x*0.25, -0.25) rectangle (1.25+\x*0.25, 0.25);
}\foreach \colour / \block in {cyan/58, red/59, green/56, yellow/57}
{
  \pgfmathtruncatemacro \from {\block}
  \pgfmathtruncatemacro \to {\block}
  \foreach \x in {\from, ..., \to}
    \draw[fill=\colour] (1+\x*0.25, -0.25) rectangle (1.25+\x*0.25, 0.25);
}

\node[fill=cyan] at (3,-1.5) {\textbf{\textit{Thread 0}}};
\node[fill=red] at (6.75,-1.5) {\textbf{\textit{Thread 1}}};
\node[fill=green] at (10.5,-1.5) {\textbf{\textit{Thread 2}}};
\node[fill=yellow] at (14.25,-1.5) {\textbf{\textit{Thread 3}}};
\end{tikzpicture}
\end{center}
\caption{Example with four threads of the three scheduling strategies offered 
by OpenMP: static, dynamic (chunk=2) and guided. Note that under the guided 
scheme chuck size is reduced exponentially.}
\label{fig:omp_scheduling}
\end{figure*}

OpenMP offers three different scheduling strategies for parallel loops: static, 
dynamic and guided \cite{openmp:spec40}. There is also a more advanced
scheduling technique, known as ``work-stealing'', which is implemented by
libraries such as Intel Cilk Plus, though it is not part of the OpenMP
specification, nor is it supported (to the best of our knowledge) by any OpenMP
implementation. In this section we will present these four options and compare
them in terms of load balance, scheduling overhead and overall efficiency.

\subsection{OpenMP static}
\label{subsect:omp_static}
Under the static scheduling scheme, the iteration space is divided into equally 
large chunks which are then assigned to threads. This can be seen in the first 
example in \Figref{fig:omp_scheduling}. Partitioning of iteration space is done 
statically at the beginning of the for-loop, so there is zero scheduling 
overhead. On the other hand, this scheme can lead to significant load 
imbalance, especially in a highly diverse loop.

\subsection{OpenMP dynamic}
\label{subsect:omp_dynamic}
Dynamic scheduling is a first approach to the problem of load imbalance. 
Instead of a static partitioning of the iteration space, chunks of work are 
assigned to threads dynamically. Once a thread finishes a chunk, it takes the 
next available from the iteration space. This is shown in the middle example in 
\Figref{fig:omp_scheduling}. Access to chunks is done via atomic updates of the 
loop counter; a thread acquiring a chunk reads the current value of the loop 
counter and increments it atomically by the chunk size.

Dynamic scheduling solves imbalance problems as threads proceed to the next 
iterations of the for-loop in a fine-grained way. As an immediate consequence, 
good load balance comes at a cost. The loop counter is updated atomically and 
this constitutes a 2-way source of overhead. The two components of overhead are 
related to instruction latency and thread competition. The time it takes to 
execute an atomic instruction can vary anywhere between a standard update in L1 
(if the thread performing the update is running on the same physical core as 
the thread which last updated the shared variable) and an update in RAM (if the 
last thread to update the shared variable is running on another socket in case 
of NUMA systems). This may not be a problem in short for-loops, but becomes 
easily a hotspot in loops with millions of iterations and little work per 
iteration (i.e. when atomic instruction latency is comparable to the loop body 
itself). Secondly, as the number of threads increases, so does the competition 
for the shared variable, leading to either (depending on the architecture) 
increased locking or increased number of failed update attempts, thus making 
atomic instruction latency even longer.

It could be argued that this overhead can be mitigated by increasing the chunk 
size, therefore lowering the number of times a thread will need to access the
loop counter. On the other hand, increasing the chunk size can introduce load
imbalance once again. Additionally, it is usually impossible to know the 
optimal chunk size at compile time and/or it can vary greatly between 
successive executions of an algorithm. Besides, relying on the chunk size for 
performance optimization puts an extra burden on the programmer.

We have found that using dynamic scheduling over guided in \PRAGMATIC can 
increase the execution time of specific algorithms by up to three times, as 
will be shown in Section \ref{sect:results}. Following that, dynamic scheduling 
was rejected as an option for that framework.

\subsection{OpenMP guided}
\label{subsect:omp_guided}
The guided scheme is an attempt to reduce dynamic scheduling overhead while 
retaining good load balance. The key difference between the two strategies is 
how chunks of work are allocated. Whereas in dynamic scheduling the chunk size 
is constant, the guided scheme starts with large chunks and the size is reduced 
exponentially as threads proceed to subsequent iterations of the for-loop. This 
can be seen in the last example of \Figref{fig:omp_scheduling}. Initial large 
chunks account for reduced atomic accesses to the loop counter while the more 
fine-grained control towards the end of the loop tries to maintain good load 
balance.

For the most part, guided scheduling works well in \PRAGMATIC, yet there are 
cases where we have observed significant load imbalance. This can happen if, 
for instance, most of the work in an irregular loop is accumulated in a few of 
the initial big chunks. In a case like that, threads processing the ``loaded'' 
chunks are busy for long while others go through the remaining ``light'' 
iterations relatively quickly and reach the end of the for-loop early, waiting 
for the busy workers to finish as well.

\subsection{Work-stealing}
\label{subsect:work_stealing}
Work-stealing (\cite{Blumofe:1994:SMC:1398518.1398998,
conf/isaac/TchiboukdjianGTRB10}) is a more sophisticated technique aiming at
balancing workload among threads while keeping scheduling overhead as low as
possible. The generic work-stealing algorithm for a set of tasks
\cite{Blumofe:1994:SMC:1398518.1398998} can be summarized as follows. Each
thread keeps a deque (double-ended queue) of tasks to execute. While the deque
is full, the thread pops workitems from the front. Once the deque is empty, the
thread becomes a thief, i.e. it chooses a victim thread randomly and steals a
chunk of workitems from the back of the victim's deque.

For a parallel for-loop with a predefined number of iterations $N$ the deques 
can simply be replaced with pairs of indices $<i_{start},i_{end}>$ 
corresponding to the range in the iteration space 
$\left[i_{start},i_{end}\right)$ which has been assigned to each thread, $0 
\leq i_{start},i_{end} < N$. In this case, every thread executes iterations by 
using $i_{start}$ as the loop counter whereas thieves steal work by 
decrementing a victim's $i_{end}$.

Accesses to those pairs of indices can lead to race conditions, so they need to 
be accessed with atomics. Following that, work-stealing for for-loops comes 
close to OpenMP's dynamic scheduling with some chunk size $>1$, with a major 
difference being that in work-stealing threads do not compete all together for 
atomic access to the same shared variable (the common loop counter); instead, 
congestion is rare and happens only if two thieves try to steal from the same 
victim.

Performance can still suffer from load imbalance and scheduling overhead when 
using work-stealing. The main drawback of the classic work-stealing algorithm 
is that thieves choose victims randomly. There is no heuristic method to 
indicate which threads are more suitable victims (i.e. have more remaining 
workload) than others. Stealing comes at a cost and picking victims with too 
little or no remaining work is inefficient, as it leads to the need for 
frequent stealing which induces some overhead. Additionally, failed attempts do 
not help balance the workload. As an example of an extreme case, a single 
thread becomes the sole remaining worker while the rest waste time trying to 
steal from each other in vain.

Mitigating the effects of random choice was our main concern when designing the 
new for-loop scheduler. We devised a low-overhead heuristic method for finding 
appropriate victims. At the same time, we tried to reduce scheduling overhead 
by eliminating the need to use atomics when accessing each thread's 
$<i_{start},i_{end}>$ pair. The following section describes in detail how the 
scheduler is implemented.

\section{Interrupt-Driven Work Sharing}
\label{sect:new_alg}
Our new scheduler differs from existing work-stealing approaches in two major 
ways. First of all, as was mentioned in Section \ref{subsect:work_stealing}, 
every worker constantly ``advertises'' its progress so that thieves can find 
suitable victims which have been left behind. Secondly, a thief does not 
actually steal work from the victim in the classic sense; instead, it 
interrupts the chosen victim by sending a POSIX signal. The signal handler 
executed by the victim encapsulates the code with which the victim decides what 
portion of its remaining workload can be given away. As it becomes apparent, 
the new scheduling algorithm is much closer to work-sharing than work-stealing, 
therefore we call it Interrupt-Driven Work-Sharing (IDWS). Nonetheless, we will 
use work-stealing terminology throughout this article.

\begin{algorithm}[h]
\caption{Parallel loop executed by each thread}
\label{alg:cooperative_loop}
\begin{algorithmic}
\For{$i = i_{start}$; $i < i_{end}$; $i \gets i+1$}
\State flush $i$ \Comment from register file to memory, so that
\State \Comment thieves can see this thread's progress
\State execute $i_{th}$ iteration
\State 
\State flush $i_{end}$ \Comment from memory to register file, as it may
\State \Comment have been modified by the signal handler
\EndFor
\end{algorithmic}
\end{algorithm}

\begin{algorithm}[h]
\caption{Work-stealing}
\label{alg:cooperative_steal}
\begin{algorithmic}
\ForAll{threads $t_n$}
\State $remaining_{t_n} \gets i_{end,t_n} - i_{start,t_n} - i_{t_n}$
\EndFor
\State let T $\gets$ $t_n$ for which $remaining_{t_n} = max$
\State send signal to victim T
\State wait for answer
\State update own $i_{start}, i_{end}$
\State execute loop chunk
\end{algorithmic}
\end{algorithm}

\begin{algorithm}[h]
\caption{Signal handling}
\label{alg:cooperative_sighandler}
\begin{algorithmic}
\State $remaining \gets i_{end} - i_{start} - i$
\If{$remaining > 1$}
\State $chunk \gets remaining/2$
\State $i_{end,thief} \gets i_{end}$
\State $i_{start,thief} \gets i_{end}-chunk$
\State $i_{end} \gets i_{end}-chunk$
\EndIf
\State send reply to thief
\end{algorithmic}
\end{algorithm}

The abstract description of this scheme can be split into three parts:

\paragraph{Loop execution (Algorithm \ref{alg:cooperative_loop})}
Every thread executes the iterations of the chunk it has acquired in the same 
way as it would using OpenMP's static scheduling scheme. Initially, the 
iteration space is divided statically into chunks of equal size and every 
thread $t_n$ is assigned one chunk. The chunk's boundaries for thread $t_n$, 
referred to as $i_{start,t_n}$ and $i_{end,t_n}$, are globally visible 
variables accessible by all threads. Compared to static scheduling, the 
important addition here is some necessary flushing of the loop counter 
$i_{t_n}$ and the loop boundary $i_{end,t_n}$. More precisely, the value of 
$i_{t_n}$ has to be written back to memory (instead of being cached in some 
register) at the beginning of every iteration so that potential thieves can 
monitor $t_n$'s progress, calculate how much work is left for $t_n$ and decide 
whether it is worth stealing from it. Similarly, the end boundary $i_{end,tn}$ 
has to read by $t_n$ from memory (instead of caching it in some register) 
before proceeding to the next iteration because $i_{end,t_n}$ might have been 
modified by the signal handler if a thief interrupted $t_n$ while the latter 
was executing an iteration of the for-loop.

\paragraph{Choosing suitable victims (Algorithm \ref{alg:cooperative_steal})}
By flushing their loop counters, threads advertise their progress so potential 
thieves can find where to steal from. When a thread becomes a thief, it 
calculates the remaining workload for all other threads by reading the 
associated values $i_{t_n}$, $i_{start,t_n}$ and $i_{end,t_n}$. This way, we 
have a heuristic method for finding which thread has the most remaining work, 
thus being a more suitable victim than others. This heuristic may not be 
optimal, but is an improvement over random choice. Once the thief has spotted 
its victim, it sends a signal and waits for an answer. The victim executes the 
signal handler and replies with the boundaries (a pair of $<i_{start}, 
i_{end}>$) of the chunk it wants to give away. Finally, the thief becomes a 
worker once again and moves on to process the newly acquired chunk.

\paragraph{Signal handler (Algorithm \ref{alg:cooperative_sighandler})}
When a victim is interrupted by the signal, control is transferred by the 
operating system to the associated handler. Inside that code, the thread 
calculates how much work it can give away (if any), replies to the thief with 
the boundaries of the donated chunk, re-adjusts the boundaries of its own chunk 
and finally returns from the signal handler.

It is clear that there are no races and no need for atomic access to any loop
variables during the stealing process, as the donor is the one who decides what
will be donated. Of course, switching from user to kernel mode to execute the
signalling system call and busy-waiting for a reply from the victim involves some
overhead; however, as will be shown in the results section, this method seems to
be more efficient than classic work stealing.

\section{C++ implementation and usage}
\label{sect:implementation}
This section describes how the IDWS scheduler is implemented and how it can 
replace existing OpenMP for-loops.

\subsection{IDWS namespace}
\label{subsect:namespace}
IDWS is a namespace encapsulating all necessary data structures and functions 
used by the new scheduler. Its declaration can be seen in Code Snippet 
\ref{code:namespace}.

\begin{code}
\begin{lstlisting}
namespace IDWS{
  struct thread_state_t;
  vector<thread_state_t> thread_state;
  void SIGhandlerUSR1(int sig);
  void IDWS_Initialize();
  void IDWS_Finalize();
};
\end{lstlisting}
\caption{IDWS namespace. It consists of initialisation and finalisation 
functions, the signal handler, the definition of \texttt{struct 
thread\_state\_t} and the vector holding all \texttt{thread\_state\_t} 
instances (one per thread).}
\label{code:namespace}
\end{code}

\begin{code}
\begin{lstlisting}
struct thread_state_t{
  size_t start;
  size_t end;
  size_t processed;
  int current_ctx;
  bool active;

  int signal_arg;
  pthread_t ptid;
  pthread_mutex_t comm_lock;
  pthread_mutex_t request_mutex;
  pthread_cond_t wait_for_answer;
};
\end{lstlisting}
\caption{thread\_state\_t struct}
\label{code:thread}
\end{code}

\paragraph{struct \texttt{IDWS::thread\_state\_t}}
The heart of IDWS is a struct named \texttt{thread\_state\_t}, which 
encapsulates all variables involved in parallel loop execution and 
work-stealing. Each participating thread has its own instance of this struct, 
which is accessible by all other threads. The struct can be seen in Code 
Snippet \ref{code:thread}.
\begin{itemize}
\item \texttt{start} and \texttt{end}: Define the current chunk boundaries
\item \texttt{processed}: Is used by a thread to advertise its progress through 
the loop
\item \texttt{current\_ctx}: IDWS loops are \texttt{nowait} loops, which means 
that a thread can proceed to the rest of the program without synchronising with 
other threads. In order to know whether two threads work inside the same loop, 
so stealing work from one another is valid, a counter \texttt{current\_ctx} is 
used, which is incremented each time a thread finishes a loop. Here we assume 
that all threads will go through all loops of the program.
\item \texttt{active}: Indicates whether the thread is inside the loop; this 
variable is used by thieves to skip immediately threads which have also become 
thieves.
\item \texttt{signal\_arg}: POSIX signals can only have two arguments, what 
signal is to sent and to whom. The victim needs to know, however, who the thief 
is, so \texttt{signal\_arg} is used by the thief to send its ID to the victim.
\item \texttt{comm\_lock}: In order to avoid needless busy-waiting by other 
thieves while one thief has already sent a signal to its victim, we use a lock 
(in form of a mutex); while this lock is held by a thief, other thieves will 
choose other victims to steal from.
\item \texttt{ptid}: POSIX ID of the thread; it is used by the thief to 
raise the signal.
\item \texttt{request\_mutex} and \texttt{wait\_for\_answer}: POSIX mutex and 
condition variables which assist the process of sending the signal and waiting 
for a reply.
\end{itemize}

Note that we need two separate mutexes and cannot use \texttt{request\_mutex} 
in place of \texttt{comm\_lock}. The former is implicitly released by the thief 
in order to enable the victim to signal the condition variable; in the 
meantime, before the victim locks \texttt{request\_mutex}, another thief might 
acquire the lock and 
destroy the process.

\paragraph{vector \texttt{IDWS::thread\_state}}
Each thread has its own instance of the \texttt{thread\_state\_t} struct. All 
instances are held in a shared vector called \texttt{thread\_state}.

\paragraph{Initialisation and finalisation}
Like MPI, IDWS needs to be initialised by calling 
\texttt{IDWS::IDWS\_Initialize()}. Threads must create their 
\texttt{thread\_state\_t} structs and push them back into the shared vector
\texttt{thread\_state}. Struct initialisation includes finding POSIX IDs and 
initialising \texttt{comm\_lock}, \texttt{request\_mutex} and 
\texttt{wait\_for\_answer}. Similarly, this data has to be destroyed at the end 
of the program, which is done by a call to \texttt{IDWS::IDWS\_Finalize()}. 
Additionally, we must register a signal handler to serve the interrupt. We have 
chosen signal \texttt{SIGUSR1} and function \texttt{IDWS::SIGhandlerUSR1} as 
the signal handler. Choice of \texttt{SIGUSR1} was arbitrary; it should be 
noted, however, that if an application uses the same signal for other purposes, 
it must re-register the original handler upon finishing with IDWS or use a 
different signal in the first place.

\paragraph{Signal handler}
A victim decided what portion of its chunk can be donated by executing the 
signal handler. The way it is done is described in Code Snippet 
\ref{code:signal_handler}. The victim first checks that the thief works in the 
same context. Then, it calculates how much work it can give away using 
\texttt{start}, \texttt{end} and \texttt{processed}, also leaving a safety 
margin due to an uncertainty regarding the true value of \texttt{processed}. In 
case of success, the victim updates both the thief's and its own \texttt{start} 
and \texttt{end} and sets \texttt{sig\_arg=1} to indicate successful donation 
(otherwise, \texttt{sig\_arg} is set to another value). Finally, the victim 
signals the condition variable to let the thief know that the signal handler is 
over.

\begin{code*}
\begin{lstlisting}
void SIGhandlerUSR1(int sig){
  int tid = omp_get_thread_num();
  int sig_thread = thread_state[tid].signal_arg; // Who sent the signal
  pthread_mutex_lock(&thread_state[tid].request_mutex);
  
  // Only share a chunk if both threads are in the same context
  if(thread_state[tid].current_ctx == thread_state[sig_thread].current_ctx){
    size_t remaining = thread_state[tid].end - thread_state[tid].start -
                       thread_state[tid].processed;
    // Leave a safety margin - we do not know if the signal was caught before,
    // after or even in the middle of updating thread_state[tid].processed.
    if(remaining > 0){
      --remaining;
      size_t chunk = remaining/2;
      thread_state[sig_thread].start = thread_state[tid].end - chunk;
      thread_state[sig_thread].end = thread_state[tid].end;
      thread_state[tid].end -= chunk;
      thread_state[tid].signal_arg = -1; // reply success
    }else thread_state[tid].signal_arg = -2; // reply failure
  }else thread_state[tid].signal_arg = -2;
  
  pthread_cond_signal(&thread_state[tid].wait_for_answer);
  pthread_mutex_unlock(&thread_state[tid].request_mutex);
}
\end{lstlisting}
\caption{Signal handler.}
\label{code:signal_handler}
\end{code*}

\subsection{Prologue and epilogue macros}
\label{subsect:macros}
The new scheduler is defined in two parts, using macros \texttt{IDWS\_prologue} 
and \texttt{IDWS\_epilogue}. These macros must surround the loop body.

\begin{code*}
\begin{lstlisting}
/* IDWS_prologue(TYPE,NAME,SIZE) starts expanding here */
  // assume tid = omp_get_thread_num();
  thread_state[tid].start = ...; thread_state[tid].end = ...;
  thread_state[tid].processed = 0; thread_state[tid].active = true;
  
  do{
    size_t __IDWS_cnt= 0;
    for(TYPE NAME=thread_state[tid].start; ; ++NAME, ++__IDWS_cnt){
      // Force flushing the progress back into memory
      *((volatile size_t *) &thread_state[tid].processed) = __IDWS_cnt;
      // Force re-laoding the end boundary from memory
      if(NAME >= *((volatile size_t *) &thread_state[tid].end)) break;
/* IDWS_prologue ends here */

      /******************************
       * loop body is executed here *
       ******************************/

/* IDWS_epilogue starts expanding here */
    } // end for
    
    thread_state[tid].active = false; // become a thief
    std::map<int,size_t> remaining;
    forall(t in active threads) // only check non-thieves
      remaining[t] = thread_state[t].end - thread_state[t].start - thread_state[t].processed;
    traverse remaining from largest to smallest;
    victim = first thread t for which pthread_mutex_trylock(&thread_state[t].comm_lock) succeeds;
    if(no victim found) break; // exit the do-while loop
    
    thread_state[victim].sig_arg = tid; // tell the victim who we are
    pthread_mutex_lock(&thread_state[victim].request_mutex);
    pthread_kill(thread_state[victim].ptid, SIGUSR1); // send signal
    pthread_cond_wait(&thread_state[victim].wait_for_answer,&thread_state[victim].request_mutex);
    pthread_mutex_unlock(&thread_state[victim].request_mutex);
    if(thread_state[victim] == -1) thread_state[tid].active = true; // become a worker again
    pthread_mutex_unlock(&thread_state[victim].comm_lock);
  } while(thread_state[tid].active = true) // end do
  
  thread_state[tid].current_context++; // proceed to next loop
/* IDWS_epilogue ends here */
\end{lstlisting}
\caption{Pseudo-code demonstrating how \texttt{IDWS\_prologue} and 
\texttt{IDWS\_epilogue} are expanded around the loop body.}
\label{code:macros}
\end{code*}

\paragraph{\texttt{IDWS\_prologue} macro}
Before entering a loop, the iteration space is split into equal chunks which 
are assigned to threads. After that, each thread begins the execution of its 
chunk. Compared to a standard for-loop, a IDWS for-loop is defined slightly 
differently. Apart from checking for the end of the loop and incrementing the 
counter after every iteration, in IDWS we must also enforce the compiler to 
flush the counter back to memory and load the updated value of $i_{end}$ from 
memory (which might have been modified by the signal handler), as indicated by 
Algorithm \ref{alg:cooperative_loop}. Flushing is done selectively for those 
two variables by casting them to volatile datatypes. Using \texttt{\#pragma omp 
flush} would flush the entire shared program state, which is not efficient. A 
pseudo-code of how the macro expands is given in Code Snippet 
\ref{code:macros}. Parameters \texttt{TYPE}, \texttt{NAME} and \texttt{SIZE} 
correspond to the datatype of the loop counter, its name and the size of the 
iteration space, respectively. In the current implementation of the new 
scheduler we assume that loops run from 0 to \texttt{SIZE} with increments of 1 
and that the loop counter is of an unsigned integral datatype.

\paragraph{\texttt{IDWS\_epilogue} macro}
After a thread finished its chunk, it becomes a thief. That means it has to 
enter the stealing process, as described in Algorithm 
\ref{alg:cooperative_steal}. The \texttt{IDWS\_epilogue} macro serves this 
purpose. The way the macro expands can be seen in Code Snippet 
\ref{code:macros}. The thief calculated for all active workers the amount of 
remaining work. Then, starting from the worker with the highest remaining 
workload, it tries to acquire the worker's \texttt{comm\_lock}. If no suitable 
worker is found, then the thief exits the IDWS loop and proceeds to the rest of 
the code. Otherwise, the thief locks the victim's mutex, sends the signal and 
waits on the victim's condition variable for an answer. The answer comes via 
\texttt{sig\_arg}. If \texttt{sig\_arg==-1}, then the victim has set the 
thief's \texttt{start} and \texttt{end} variables, so the thief becomes a 
worker again. If any other answer has been sent back, then the thief exits the 
IDWS loop. It is important to note that a memory fence is necessary on the
thief's side between setting the victim's signal argument \texttt{sig\_arg} and
raising the signal, so that we make sure that the victim will see the correct
value of \texttt{sig\_arg}. Locking the victim's mutex before sending the signal
works as an implicit memory fence.

\subsection{OpenMP to IDWS}
\label{subsect:conversion}
\begin{code}[h]
\begin{lstlisting}
#include <omp.h>
...
int main(){
  ...
  #pragma omp parallel
  {
    ...
    #pragma omp for
    for(TYPE VAR=0; VAR<SIZE; ++VAR){
      do_something(VAR);
    }
    ...
  }
  ...
}
\end{lstlisting}
\caption{Initial OpenMP for-loop. The loop must be inside an OMP parallel 
region.}
\label{code:inital_code}
\end{code}

\begin{code}[h]
\begin{lstlisting}
#include <omp.h>
#include "IDWS.h"
...
int main(){
  ...
  IDWS::IDWS_Initialize();
  int nthreads = omp_get_max_threads();
  ...
  #pragma omp parallel
  {
    int tid = omp_get_thread_num();
    ...
    IDWS_prologue(TYPE, VAR, SIZE)
      do_something(VAR);
    IDWS_epilogue
    ...
  }
  ...
  IDWS::IDWS_Finalize();
}
\end{lstlisting}
\caption{Transformed code showing what has to be added/modified in order to use 
the new scheduler instead of a standard OpenMP scheduling strategy.}
\label{code:transformed_code}
\end{code}

The new scheduler can be used directly with virtually any C++ OpenMP 
application written for any POSIX-compliant operating system (provided that the 
compiler used implements OpenMP upon pthreads). A prerequisite for converting 
an OpenMP loop to a IDWS one is that the former is written as shown in Code 
Snippet \ref{code:inital_code}, i.e. the loop must be inside an \texttt{omp 
parallel} region. Conversion to IDWS loops is shown in Code Snippet 
\ref{code:transformed_code}. The user needs to include header file ``IDWS.h'' 
which can be downloaded from \PRAGMATIC's page  on 
Launchpad\footnote{\url{https://code.launchpad.net/~gr409/pragmatic/IDWS}}. 
This header file defines the IDWS namespace and the prologue and epilogue 
macros.

Compared to the initial version, we need to define:
\begin{itemize}
\item \texttt{int nthreads=omp\_get\_max\_threads()}: shared variable outside 
the parallel region
\item \texttt{int tid=omp\_get\_thread\_num()}: thread-private variable inside 
the parallel region
\end{itemize}
, remove the \texttt{\#pragma omp for} directive and the for-loop declaration 
and, finally, surround the loop-body with the \texttt{IDWS\_prologue} and 
\texttt{IDWS\_epilogue} macros.

\section{Experimental Results}
\label{sect:results}
\begin{table*}[h]
\begin{center}
\begin{tabular}[c]{|l|c|c|c|c|c||c|c|c|c|}
\cline{2-10}
\multicolumn{1}{c|}{}	& \multicolumn{5}{|c||}{Synthetic benchmark}		& \multicolumn{4}{c|}{\PRAGMATIC kernels}	\\ \cline{2-10}
\multicolumn{1}{c|}{}	& Regular	& Random	& Dense End	& Dense Begin	& Periodic	& Coarsen	& Refine	& Swap	& Smooth \\ \hline
IDWS					& 5.85		& 7.48		& 4.14		& 3.80			& 1.13		& 12.4		& 7.29		& 19.9	& 11.0   \\ \hline
OMP static				& 5.84		& 7.51		& 15.7		& 15.0			& 1.13		& 18.6		& 7.87		& 20.4	& 12.1   \\ \hline
OMP static,1			& 5.92		& 7.60		& 4.32		& 3.91			& 14.5		& 15.4		& 8.45		& 22.5	& 12.4   \\ \hline
OMP dynamic				& 22.4		& 27.0		& 21.4		& 20.5			& 9.72		& 67.4		& 31.4		& 99.7	& 17.9   \\ \hline
OMP guided				& 5.82		& 7.46		& 4.27		& 7.08			& 1.12		& 12.1		& 6.88		& 19.5	& 11.1   \\ \hline
Cilk+					& 6.12		& 7.76		& 4.35		& 4.00			& 1.20		& -			& -			& -		& -      \\ \hline
\end{tabular}
\caption{Execution time in seconds for each benchmark using the 6 different 
scheduling strategies on a dual-socket Intel Xeon E5-2650 (Sandy Bridge, 
2.00GHz, 8 physical cores per socket, 16 hyperthreads per socket, 32 threads in 
total).}
\label{tab:performance_cx1}
\end{center}
\end{table*}

\begin{table*}[h]
\begin{center}
\begin{tabular}[c]{|l|c|c|c|c|c||c|c|c|c|}
\cline{2-10}
\multicolumn{1}{c|}{}	& \multicolumn{5}{|c||}{Synthetic benchmark}		& \multicolumn{4}{c|}{\PRAGMATIC kernels}	\\ \cline{2-10}
\multicolumn{1}{c|}{}	& Regular	& Random	& Dense End	& Dense Begin	& Periodic	& Coarsen	& Refine	& Swap	& Smooth \\ \hline
IDWS					& 8.24		& 10.6		& 5.86		& 5.37			& 2.86		& 17.5		& 7.06		& 24.5	& 17.3   \\ \hline
OMP static				& 8.25		& 10.6		& 21.2		& 22.7			& 2.86		& 29.3		& 8.13		& 26.8	& 18.3   \\ \hline
OMP static,1			& 8.37		& 10.7		& 6.01		& 5.70			& 22.2		& 23.0		& 9.55		& 30.7	& 19.1   \\ \hline
OMP dynamic				& 26.9		& 34.6		& 22.8		& 21.7			& 9.43		& 63.7		& 26.3		& 91.8	& 22.7   \\ \hline
OMP guided				& 8.23		& 10.5		& 6.07		& 9.81			& 2.84		& 17.6		& 7.08		& 24.3	& 17.3   \\ \hline
Cilk+					& 8.38		& 10.7		& 5.96		& 5.47			& 2.92		& -			& -			& -		& -      \\ \hline
\end{tabular}
\caption{Execution time in seconds for each benchmark using the 6 different 
scheduling strategies on a dual-socket Intel Xeon E5-2643 (Sandy Bridge, 
3.30GHz, 4 physical cores per socket, 8 hyperthreads per socket, 16 threads in 
total).}
\label{tab:performance_badger}
\end{center}
\end{table*}

\begin{table*}[h]
\begin{center}
\begin{tabular}[c]{|l|c|c|c|c|c||c|c|c|c|}
\cline{2-10}
\multicolumn{1}{c|}{}	& \multicolumn{5}{|c||}{Synthetic benchmark}		& \multicolumn{4}{c|}{\PRAGMATIC kernels}	\\ \cline{2-10}
\multicolumn{1}{c|}{}	& Regular	& Random	& Dense End	& Dense Begin	& Periodic	& Coarsen	& Refine	& Swap	& Smooth \\ \hline
IDWS					& 11.0		& 19.6		& 9.60		& 9.02			& 0.97		& 30.7		& 17.2		& 86.7	& 26.9   \\ \hline
OMP static				& 12.3		& 22.3		& 49.0		& 54.2			& 1.06		& 34.7		& 19.7		& 79.1	& 27.7   \\ \hline
OMP static,1			& 12.6		& 22.9		& 11.2		& 10.6			& 48.3		& 35.6		& 21.2		& 122	& 26.2   \\ \hline
OMP dynamic				& 40.1		& 31.5		& 25.7		& 25.1			& 15.2		& 129		& 59.6		& 234	& 29.4   \\ \hline
OMP guided				& 10.8		& 19.6		& 10.3		& 23.5			& 0.92		& 29.9		& 15.6		& 85.3	& 24.0   \\ \hline
Cilk+					& 11.3		& 19.9		& 10.1		& 9.59			& 1.05		& -			& -			& -		& -      \\ \hline
\end{tabular}
\caption{Execution time in seconds for each benchmark using the 6 different 
scheduling strategies on Xeon Phi (1.2GHz, 61 physical cores, 2 hyperthreads 
per core, 122 threads in total).}
\label{tab:performance_phi_120}
\end{center}
\end{table*}

\begin{table*}[h]
\begin{center}
\begin{tabular}[c]{|l|c|c|c|c|c||c|c|c|c|}
\cline{2-10}
\multicolumn{1}{c|}{}	& \multicolumn{5}{|c||}{Synthetic benchmark}		& \multicolumn{4}{c|}{\PRAGMATIC kernels}	\\ \cline{2-10}
\multicolumn{1}{c|}{}	& Regular	& Random	& Dense End	& Dense Begin	& Periodic	& Coarsen	& Refine	& Swap	& Smooth \\ \hline
IDWS					& 7.46		& 15.9		& 7.46		& 7.06			& 0.56		& 34.2		& 19.9		& 177	& 25.9   \\ \hline
OMP static				& 8.13		& 16.5		& 27.0		& 27.1			& 0.51		& 29.1		& 21.3		& 174	& 19.4   \\ \hline
OMP static,1			& 7.65		& 16.0		& 7.63		& 7.24			& 24.7		& 27.9		& 20.0		& 202	& 19.3   \\ \hline
OMP dynamic				& 17.3		& 19.8		& 13.9		& 13.6			& 7.68		& 81.4		& 38.4		& 247	& 24.4   \\ \hline
OMP guided				& 7.27		& 15.7		& 8.11		& 19.6			& 0.52		& 96.1		& 35.1		& 275	& 46.6   \\ \hline
Cilk+					& 8.03		& 16.3		& 8.31		& 7.97			& 0.63		& -			& -			& -		& -      \\ \hline
\end{tabular}
\caption{Execution time in seconds for each benchmark using the 6 different 
scheduling strategies on Xeon Phi (1.2GHz, 61 physical cores, 4 hyperthreads 
per core, 244 threads in total).}
\label{tab:performance_phi_240}
\end{center}
\end{table*}

\begin{figure}[h]
\begin{center}
\includegraphics[width=1.0\linewidth]{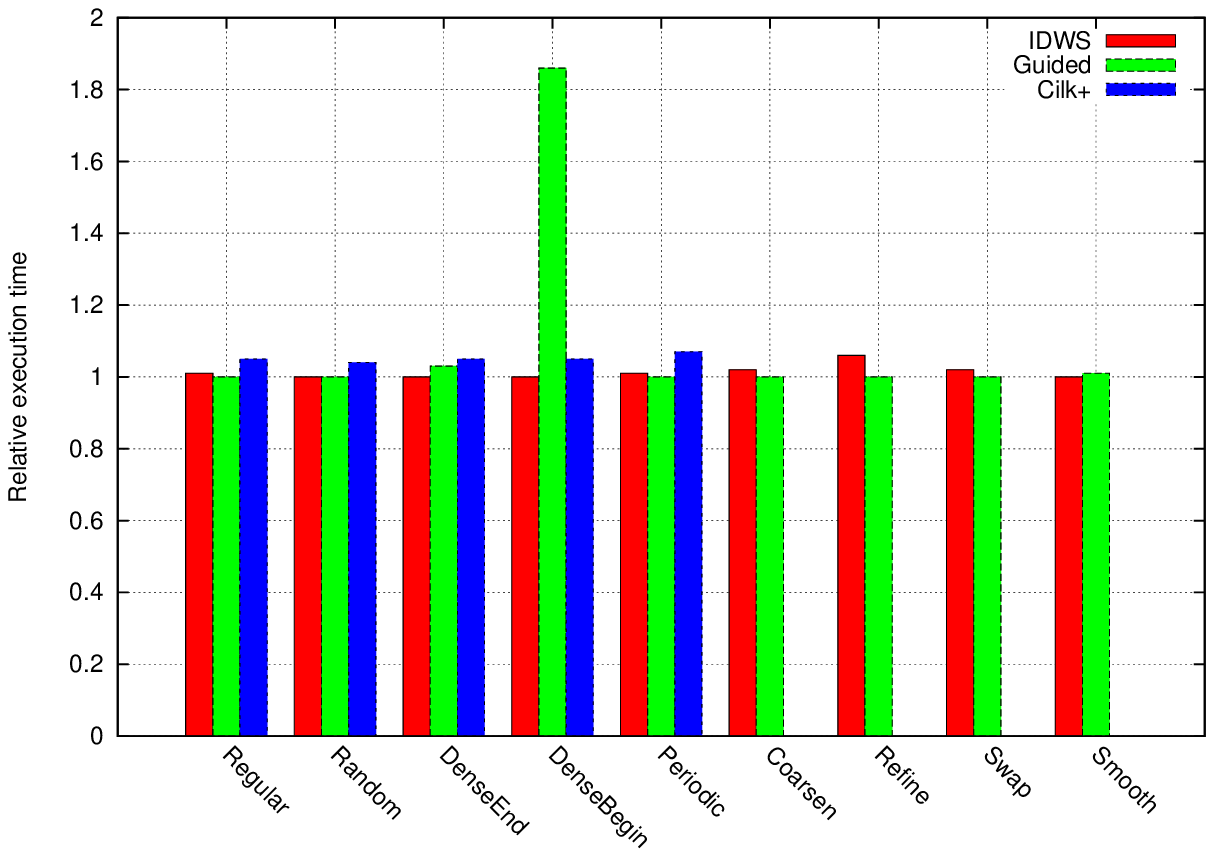}
\caption{Relative execution time (lower is better) between IDWS, OpenMP guided
and Cilk Plus on the Intel Xeon E5-2650 system. For each benchmark, the fastest
scheduling strategy is taken as reference (scoring 1.0 on the y-axis).}
\label{fig:cx1}
\end{center}
\end{figure}

\begin{figure}[h]
\begin{center}
\includegraphics[width=1.0\linewidth]{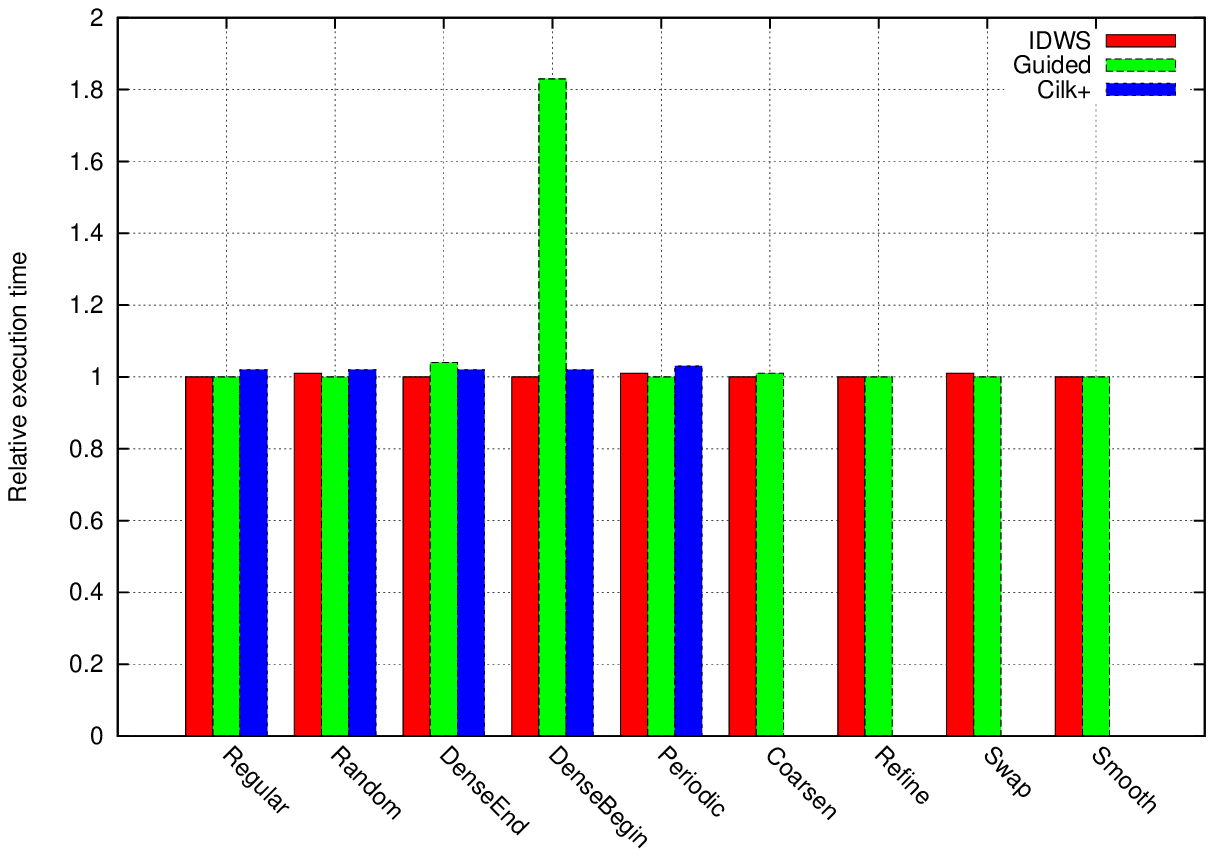}
\caption{Relative execution time (lower is better) between IDWS, OpenMP guided
and Cilk Plus on the Intel Xeon E5-2643 system. For each benchmark, the fastest
scheduling strategy is taken as reference (scoring 1.0 on the y-axis).}
\label{fig:badger}
\end{center}
\end{figure}

\begin{figure}[h]
\begin{center}
\includegraphics[width=1.0\linewidth]{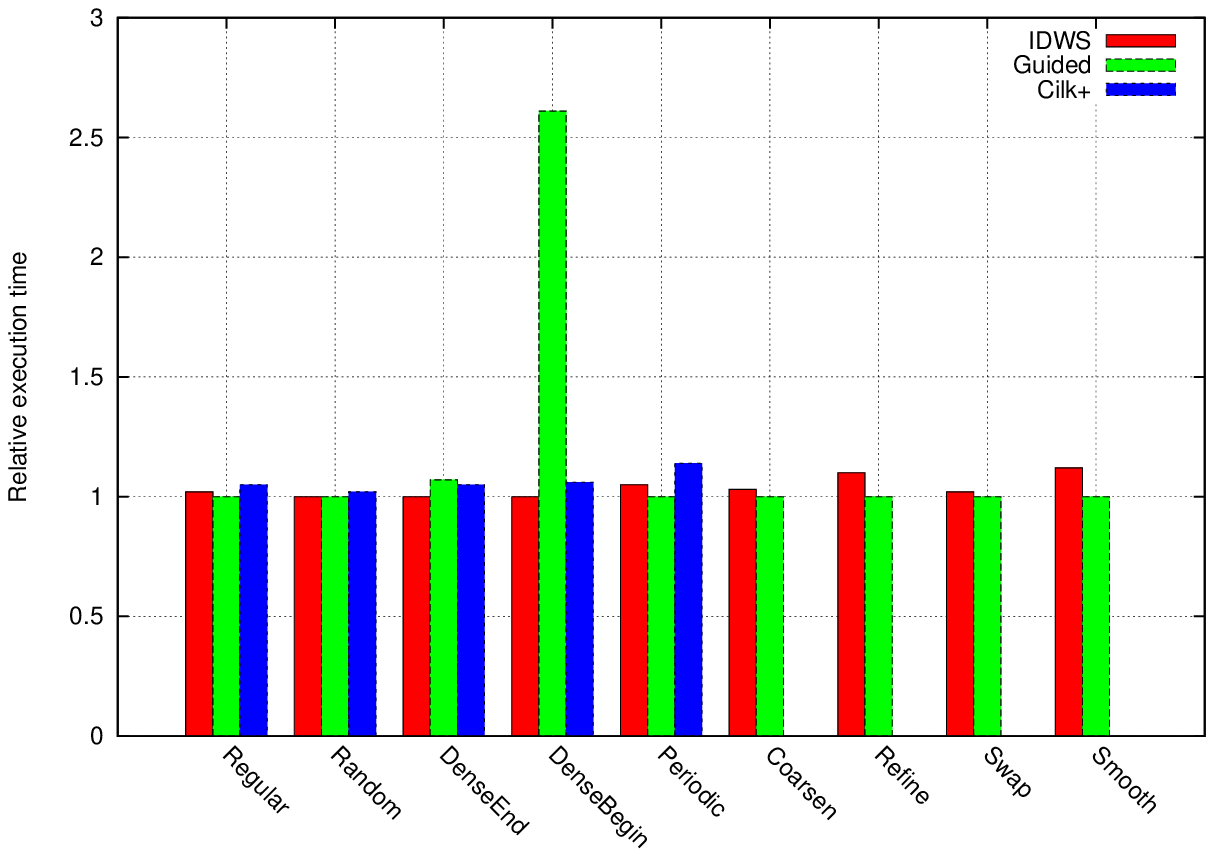}
\caption{Relative execution time (lower is better) between IDWS, OpenMP guided
and Cilk Plus on Intel Xeon Phi using 122 threads. For each benchmark, the
fastest scheduling strategy is taken as reference (scoring 1.0 on the y-axis).}
\label{fig:phi_120}
\end{center}
\end{figure}

\begin{figure}[h]
\begin{center}
\includegraphics[width=1.0\linewidth]{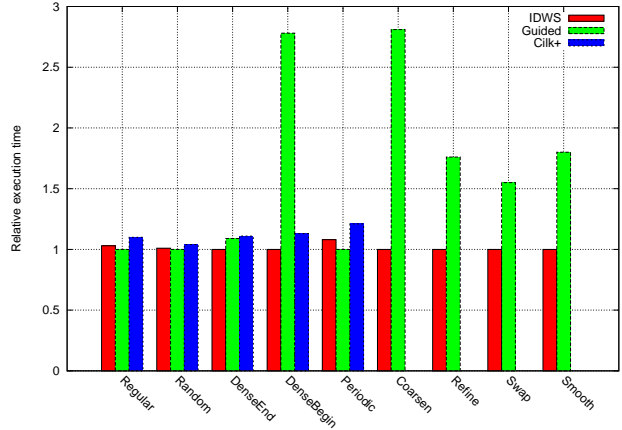}
\caption{Relative execution time (lower is better) between IDWS, OpenMP guided
and Cilk Plus on Intel Xeon Phi using 244 threads. For each benchmark, the
fastest scheduling strategy is taken as reference (scoring 1.0 on the y-axis).}
\label{fig:phi_240}
\end{center}
\end{figure}

In order to measure the performance of our new scheduler, we ran a series of
tests using both synthetic benchmarks and real kernels from the \PRAGMATIC
framework. We used three systems: a dual-socket Intel Xeon E5-2650 (Sandy
Bridge, 2.00GHz, 8 physical cores per socket, 2 hyperthreads per core, 32
threads in total), a dual-socket Intel Xeon E5-2643 (Sandy Bridge, 3.30GHz, 4
physical cores per socket, 2 hyperthreads per core, 16 threads in total) and an
Intel Xeon Phi 0b/01 board (1.2GHz, 61 physical cores, 4 hyperthreads per core,
244 threads in total). The two Xeon systems run Red Hat Enterprise Linux Server
release 6.4 (Santiago). Both versions of the code (intel64 and mic) were
compiled with Intel Composer XE 2013 SP1 using the -O3 optimisation flag. The
benchmarks were run using Intel's thread-core affinity support with the maximum
number of available threads on each platform. Additionally, we ran a second
series of benchmarks on Xeon Phi using half the available number of threads and
more specifically using all 61 cores with 2 hyperthreads per core. We did so
because we have observed that for some codes Xeon Phi performs best when using
this threading configuration.

The synthetic benchmark was designed to be compute-bound with minimal memory
traffic and no thread synchronization. Our purpose is to show how the different
scheduling strategies compare to each other in terms of achievable load balance
and incurred scheduling overhead without being affected by other factors (such
as memory bandwidth, data locality etc.). The synthetic benchmark uses an array
\mbox{\texttt{int states[16M]}}, which is populated with values in the range
$[0..3]$. Then, the parallel loop iterates over this array. For each element i,
$i \in [0..16M)$, the kernel performs a different amount of work according to
the value of \texttt{states[i]}. If \texttt{states[i]==0}, nothing is done. If
\texttt{states[i]==1}, the kernel computes $\sin()$ values of \texttt{i} and
powers of \texttt{i}. If \texttt{states[i]==2}, the kernel additionally computes
$\cos()$ values of \texttt{i} and its powers. Finally, if \texttt{states[i]==3},
the kernel additionally computes some $\sinh()$ values.

Array \texttt{states} is populated five times with different distributions of
workload and total amount of work. Each population has been given a name:
\begin{itemize}
\item Regular: All elements of \texttt{states} are set equal to 2. This is a
distribution corresponding to a regular loop which does the exact same thing in
every iteration.
\item Random: \texttt{states} is populated with random values following a uniform
distribution. This sub-benchmark corresponds to real-life distributions in
problems like graph coloring or the swap and smooth kernels in \PRAGMATIC.
\item Dense End: Most of the workload is accumulated towards the end of the
iteration space, where \texttt{states[i]=3}, while the beginning is populated
with a uniform mixture of values $[0..3)$. The rest of the iteration space is
set to 0, i.e. no work. This is a distribution closely related to the refinement
kernel in \PRAGMATIC.
\item Dense Start: Mirrored distribution of Dense End. Closely related to the
\PRAGMATIC's coarsening kernel. This is an example of workload distribution for
which OpenMP guided scheduling is a bad choice.
\item Periodic: There is a repeating pattern of states throughout the iteration
space. It is particularly bad for static scheduling with interleaved allocations
of iterations (i.e. with some chunk size).
\end{itemize}

Apart from the synthetic benchmark, we also ran \PRAGMATIC using the various
scheduling options in order to see how each strategy performs in a real-life
scenario, where compute capacity is not the only performance-limiting factor. It
should be noted that \PRAGMATIC is build upon OpenMP, so there are no results
for Cilk+ in this case.

Table \ref{tab:performance_cx1}, Table \ref{tab:performance_badger} and
Tables \ref{tab:performance_phi_120} \& \ref{tab:performance_phi_240} show the
execution time on the three platforms, respectively, using six scheduling
strategies for each distribution of the synthetic benchmark and the four
\PRAGMATIC kernels. The strategy named ``OMP static,1'' is static scheduling
with chunk size equal to 1. As can be seen, IDWS is either the fastest
scheduling option or very close to the fastest for each benchmark-platform
combination. Additionally, it clearly outperforms Cilk Plus, with the
performance gap becoming wider as the number of threads increases and Cilk's
design to pick victims in a random fashion becomes inefficient. Those results
are a good indication that IDWS is future-proof and ready for the thousand-core
era.

Regarding \PRAGMATIC, IDWS's major competitor seems to be OpenMP's guided
scheduling. Despite not being very suitable for certain kernels (coarsening)
theoretically, in practice it performs just as well as IDWS. A notable exception
is the 244-thread case on Xeon Phi, where guided scheduling is the worst choice
among the available options.

A comparison of relative performance between the three major competitors (IDWS,
OpenMP guided and Cilk Plus) is shown in Figure \ref{fig:cx1} (Intel Xeon
E5-2650 system), Figure \ref{fig:badger} (Intel Xeon E5-2643 system), Figure
\ref{fig:phi_120} (Intel Xeon Phi with 122 threads) and Figure
\ref{fig:phi_240} (Intel Xeon Phi with 244 threads). For each benchmark, we
compare the relative execution time between IDWS, OpenMP guided and Cilk Plus
(for \PRAGMATIC kernels there is only IDWS vs OpenMP guided comparison).
Reference execution time per benchmark, i.e. the one which corresponds to 1.0 on
the y-axis, is execution time of the fastest scheduler.

\section{Conclusions and Future Work}
\label{sect:conclusions}
We have presented an Interrupt-Driven Work-Sharing for-loop scheduler which is
based on work-stealing principles and tries to address major problem of the
original work-stealing algorithm: random choice of victims. The first
implementation of IDWS works very efficiently, outperforming Intel Cilk Plus,
while being from slightly slower to considerably faster than the best (per
kernel) OpenMP scheduling strategy. These results indicate that IDWS could
become the universal default scheduler for OpenMP for-loops, freeing the
programmer from tricky and disruptive management of load balance.

Two main points of focus for further work should be data locality and
efficiency of work-sharing. Work on locality issues has been published by
several groups (\cite{Olivier:2012:OTS:2237840.2237846, 5470425, 4536188,
Acar:locality}), whereas Adnan and Sato have presented interesting ideas on
efficient work-stealing strategies \cite{6008879}, some of which could be
applicable to our work-sharing scheduler.

\bibliographystyle{abbrv}
\bibliography{paper.bib}

\end{document}